\documentclass[12pt,a4paper]{article}
\pdfoutput=1
\usepackage{amsmath,amssymb,exscale,cancel}
\usepackage{graphicx}
\usepackage{epsfig}
\usepackage{multicol}
\usepackage{color}
\usepackage{mathrsfs}
\usepackage{blindtext}
 \usepackage{fancyhdr}
\usepackage{hyperref}
\usepackage{cite}
\usepackage{mathtools}

\usepackage{floatrow}

\usepackage{graphicx}
\usepackage{sidecap}

\usepackage[latin1]{inputenc} 
\usepackage{multicol}  
 
 \usepackage{titlesec}
 
 \usepackage{rotating,slashed,xcolor,amsfonts,expdlist,charter}
 
\numberwithin{equation}{section}

\usepackage{xcolor}
\usepackage{sectsty}

\usepackage{mdframed}
\usepackage{titletoc}

\definecolor{secnum}{RGB}{13,151,225}
\definecolor{ptcbackground}{RGB}{212,237,252}
\definecolor{ptctitle}{RGB}{0,177,235}

\titlecontents{lsection}
  [5.8em]{\sffamily}
  {\color{secnum}\contentslabel{2.3em}\normalcolor}{}
  {\titlerule*[1000pc]{.}\contentspage\\\hspace*{-5.8em}\vspace*{5pt}%
    \color{white}\rule{\dimexpr\textwidth-15.5pt\relax}{1pt}}

\usepackage{hyperref}
\hypersetup{colorlinks,bookmarksopen,bookmarksnumbered,citecolor=rossos,
linkcolor=redy,pdfstartview=FitH,urlcolor=blus}
\usepackage{slashed}

\definecolor{blus}{cmyk}{1,1,0,0.1}
\definecolor{verdes}{cmyk}{0.99,0,0.59,0.65}
\definecolor{rossos}{cmyk}{0,1,1,0.55}
\definecolor{redy}{cmyk}{0,1,1,0.7}
\definecolor{greeny}{cmyk}{0.99,0,0.59,0.98}
\definecolor{green-go}{cmyk}{0.79,0,0.59,0.5}

\usepackage{titlesec}

\newcommand{\beq}{\begin{equation}}
\newcommand{\eeq}{\end{equation}}

\newcommand{\pc}{{\rm pc}}
\newcommand{\Mpc}{{\rm Mpc}}
\newcommand{\Msun}{{M_{\odot}}}

\newcommand{\tmtextbf}[1]{{\bfseries{#1}}}
\newcommand{\tmtextrm}[1]{{\rmfamily{#1}}}
\newcommand{\bp}{\bar M_P}

\def\be{\begin{equation}}
\def\ee{\end{equation}}
\def\ba{\begin{array} }
\def\bac{\begin{array} {c}}
\def\bacc{\begin{array} {cc}}
\def\baccc{\begin{array} {ccc}}
\def\bacccc{\begin{array} {cccc}}
\def\ea{\end{array}}
\newcommand{\bea}{\begin{equation}\begin{aligned}}
\newcommand{\eea}{\end{aligned}\end{equation}}

\definecolor{red}{rgb}{1,0,0}

\def\psl{\hbox{\hbox{${p}$}}\kern-1.9mm{\hbox{${/}$}}}
\def\dsl{\hbox{\hbox{${\partial}$}}\kern-2.2mm{\hbox{${/}$}}}
\def\Dsl{\hbox{\hbox{${D}$}}\kern-2.6mm{\hbox{${/}$}}}

\newcommand{\gappeq}{{\rlap{{\raise}.5ex\text{\ensuremath{>}}}{{\lower}.5ex\text{\ensuremath{\sim}}}}}
\newcommand{\lappeq}{{\rlap{{\raise}.5ex\text{\ensuremath{<}}}{{\lower}.5ex\text{\ensuremath{\sim}}}}}

\newcommand{\I}{\tmtextrm{1{\kern}-.24em l}}

\newcommand{\td}{\mathrm{d}}

\begin{document}
\topmargin -1.0cm
\oddsidemargin 0.9cm
\evensidemargin -0.5cm

{\vspace{-1cm}}
\begin{center}

\vspace{-1cm}

 {\Huge \tmtextbf{ \color{blus} 
Horizonless ultracompact objects and dark matter in quadratic gravity}} {\vspace{.5cm}}\\
 
\vspace{0.9cm}

{\large  {\bf Alberto Salvio}$^a$ and  {\bf Hardi Veerm\"ae}$^b$
\vspace{.3cm}

{\em  

\vspace{.4cm}
${}^a$ Physics Department, University of Rome and INFN Tor Vergata, Italy\\ 
${}^b$ National Institute of Chemical Physics and Biophysics,  R\"avala 10, 10143 Tallinn, Estonia
\vspace{0.4cm}

\vspace{0.2cm}

 \vspace{1.5cm}
}}

\noindent

\end{center}

\begin{center}
{\bf \large Abstract}
\end{center}

\noindent We show that in quadratic gravity sufficiently light objects must be horizonless and construct explicit analytic examples of horizonless ultracompact objects (UCOs), which are more compact than Schwarzschild black holes. Due to the quadratic terms, gravity becomes soft and eventually vanishes in the high-energy limit leading to a ``linearization mechanism": light objects can be described by the linearized theory when their Schwarzschild radius is smaller than the Compton wavelength of the new gravitational degrees of freedom. As a result, we can analytically describe UCOs with a mass-to-radius ratio higher than for a Schwarzschild black hole. The corresponding spacetime is regular everywhere. We show that the Ostrogradsky instabilities can be avoided and discuss the relation with the Higgs vacuum metastability. Due to the lack of a horizon, light UCOs do not evaporate. Therefore, they may play the role of dark matter. We briefly discuss their phenomenology.
\vspace{0.9cm}

%

\vspace{0.5cm}

\vspace{3cm}

\noindent Emails: ${}^a$alberto.salvio@roma2.infn.it, ${}^b$hardi.veermae@cern.ch

\vspace{0.cm}
%
\noindent 


\vspace{-.5cm}

\noindent 
 \newpage

\vspace{0.2cm}

\tableofcontents

\section{Introduction}

The discovery of gravitational waves, whose production is consistent with  coalescing black holes (BHs)~\cite{Abbott:2016blz} in Einstein's general relativity (GR) and neutron stars~\cite{GBM:2017lvd}, has naturally reinforced the interest in ultracompact objects (UCOs) in astrophysics and cosmology (for a recent review see e.g.~\cite{Cardoso:2019rvt}).

In GR, a sufficiently high mass-to-radius ratio implies that all matter configurations are enclosed within their event horizon and are thus BHs. The Buchdahl theorem~\cite{Buchdahl:1959zz} allows  smaller mass-to-radius ratios for stable perfect fluid stars, implying that more compact objects will collapse to a BH. \footnote{However, note that stars approaching the compactness of a BH in GR can be obtained for certain anisotropic equations of state (see e.g.~\cite{Raposo:2018rjn}).} 
On the other hand, we know that GR, although very successful in describing the observed gravitational interactions, can at most be valid up to the Planck scale. As a result a BH cannot be described everywhere by GR because of the singularity at its center: in a neighborhood of this point the curvature of spacetime is much bigger than the Plank scale. A UV completion is, therefore, needed to fully describe UCOs. 

The finite range of validity of GR manifests itself through the non-renormalizability of the theory: an infinite number of experimental inputs must be provided to fix the predictions at energies around the Planck scale. This lack of predictiveness may be avoided by adding to the action all local terms quadratic in the curvature (with dimensionless parameters): the resulting theory, known as quadratic gravity, is renormalizable as first pointed out in~\cite{Weinberg:1974tw,Deser:1975nv}  and formally proved in~\cite{Stelle:1976gc} (see also~\cite{Barvinsky:2017zlx} for a recent discussion). 

In Ref.~\cite{Salvio:2017qkx} it was also showed that quadratic gravity can be UV complete if coupled to an asymptotically free (or, more generally, asymptotically safe) matter sector: in this case the theory can flow to conformal gravity (a theory whose pure gravitational Lagrangian is only given by the squared of the Weyl tensor) in the infinite energy limit. Furthermore, quadratic gravity can screen the gravitational contributions to the Higgs mass (solving the part of the hierarchy problem due to gravity) when the coefficient of the Weyl-squared term is large enough~\cite{Salvio:2014soa,Salvio:2017qkx}. This occurs because  such term acts, loosely speaking, as an anti-graviton and thus softens the gravitational interactions above a certain scale between the electroweak (EW) and the Planck scales. Quadratic gravity is, therefore, a concrete realization of the softened gravity scenario discussed in Refs.\cite{Giudice:2014tma,Salvio:2016vxi}. This special behavior is due to the presence of higher derivatives  in the action\footnote{For this reason quadratic gravity is sometimes called higher-derivative gravity.}, which make the graviton propagator decrease as the fourth inverse power of the momentum in the UV.

Moreover, quadratic gravity is consistent with the observational constraints regarding the early universe and predicts interesting new effects potentially within the reach of future observations~\cite{Kannike:2015apa,Salvio:2017xul,Salvio:2019ewf,Salvio:2019wcp}: these include an isocurvature mode of gravitational origin \cite{Salvio:2017xul,Salvio:2019ewf,Salvio:2019wcp}. We also note that, because quadratic gravity can flow to conformal gravity in the UV and the Friedmann-Robertson-Walker metric is conformally flat, the initial-time cosmological singularity of GR can  be avoided.

Studying UCOs in quadratic gravity is thus relevant. It has been found through numerical calculations that the coupling of quadratic gravity to ordinary matter only generates horizonless solutions (at least for spherically symmetric and asymptotically flat metrics)~\cite{Lu:2015psa}.  Such solutions can mimic the observed properties of astrophysical BHs~\cite{Holdom:2002xy,Holdom:2016nek,Ren:2019jft,Holdom:2019ouz}. Moreover, it is important to note that, although the Schwarzschild metric is a solution of the vacuum equations~\cite{Lu:2015cqa,Lu:2015psa} of quadratic gravity, real-world objects cannot collapse to a point, due to the softening of gravity at short distances, and should, therefore, form through the presence of some matter source.

Here we analytically show that horizons do not form in quadratic gravity at least for sufficiently light objects, that is, for objects with Schwarzschild radii smaller than the inverse masses $1/M_2$ and $1/M_0$ of the new gravitational degrees of freedom, the massive spin-2 field and the massive scalar corresponding to the $R^2$ term in the Lagrangian (where $R$ is the Ricci scalar).\footnote{This result was briefly conjectured in~\cite{Salvio:2014soa,Salvio:2018kwh}.}  A proof is possible in the linear regime, because gravitational interactions, unlike in GR, are softened for lengths much shorter than $1/M_2$ and $1/M_0$, such that gravity can be described well by its leading contribution in the weak-field expansion around flat spacetime. We refer to this effect as the ``linearization mechanism". We can, therefore, {\it analytically} find UCOs by applying this weak-field expansion. By construction, the corresponding spacetime is regular everywhere.

Due to the lack of horizons, light UCOs do not evaporate. This has several phenomenological implications. First, unlike light BHs in GR, such objects can be stable and thus serve as possible candidates for dark matter (DM) as they form, e.g., via the collapse of large primordial fluctuations~\cite{Misner:1974qy,Carr:1974nx}. Heavier UCOs, if they possess a horizon (if they are BH), will evaporate, but have to stop doing so after they lose most of their mass and enter the softened gravity regime, leaving a remnant.  The idea of BH remnants as DM is not new~\cite{MacGibbon:1987my} and the existence of remnants due to higher-order terms was suggested in Ref.~\cite{Barrow:1992hq}, although no concrete realizations of such objects were proposed so far. Softening of gravity implies that the evaporation history of BHs must thus be changed and this can affect the allowed mass window of heavier DM candidates~\cite{Raidal:2018eoo}. However, since our analysis is restricted to solutions valid in the linear regime, the quantification of these modifications for general UCOs is not explored in this paper. 

A theory of gravity valid down to arbitrarily small distances should also shed light on the physics of such objects even at finite distances from its center. For example, a natural expectation is that a solution of the  information puzzle affecting objects with horizons could be found. Regarding our paper, this is not necessary due to the existence of remnants suggested by our results (see e.g. Refs.~\cite{Giddings:1994qt,Susskind:1995da} for issues with a mere  remnant  solution of the information paradox), but might be due to modifications of the Hawking radiation and its connection with the source of the UCO, when it possesses a horizon. We leave the quantitative study of Hawking radiation of BHs in quadratic gravity to future work (see, however, Ref.~\cite{Konoplya:2019ppy} for a related work).

Another interesting advantage of the absence of micro BHs regards the metastability of the EW vacuum. Indeed, it was found~\cite{Burda:2015isa,Burda:2016mou,Tetradis:2016vqb} that micro BHs endanger the sub-Planckian high-energy consistency of the Standard Model if their Schwarzschild radii are smaller than the instability scale of the Standard Model, which is roughly given by the point of maximum of the Higgs effective potential. 

It is a well-known fact that, due to the Ostrogradsky theorem, theories whose Lagrangian depends non-linearly on second (or higher) derivatives of canonical coordinates feature an unbounded classical Hamiltonian\cite{Ostrogradsky:1850fid} (for recent reviews see e.g.~\cite{Woodard:2015zca,Salvio:2018crh}). However, as shown in Refs.~\cite{Pavsic:2013noa,Pavsic:2016ykq,Salvio:2019ewf}, this does not necessarily imply that the relevant solutions are unstable. The way the Ostrogradsky theorem manifests itself in quadratic gravity is through the presence of a spin-2 field, whose kinetic term contributes negatively to the Hamiltonian. Nevertheless, the fact that this spin-2 field has a positive mass squared $M_2^2>0$ and a very small coupling $f_2\ll 1$ leads to an island of stability for the theory\cite{Salvio:2019ewf}: the possible Ostrogradsky instabilities are avoided if the energies in the gravitational and matter sectors satisfy certain bounds. In practice this happens thanks to the presence of a sort of energy barrier, which prevents the theory from decaying. The above mentioned island is large enough to accommodate realistic cosmological solutions\cite{Salvio:2019ewf}.

A natural worry is that an instability might show up in the quantized theory through tunneling across the barrier. To circumvent this issue several non-standard quantization schemes have been proposed. In order to render the quantum Hamiltonian bounded from below, the Hilbert space is endowed with an indefinite metric, with respect to which the canonical coordinates and conjugate momenta are self-adjoint~\cite{Pais:1950za,Stelle:1976gc,Salvio:2015gsi} (see also~\cite{Bender:2007wu,Bender:2008gh}). Such approaches are mathematically equivalent to canonically quantizing the complexified classical theory  -- this procedure yields, by construction, positive norm quantum theories~\cite{Raidal:2016wop}. Positive transition probabilities that sum up to one, i.e. satisfy unitarity, can more generally be obtained by \ replacing the indefinite metric by a suitable positively-definite one when applying the Born rule~\cite{Mostafazadeh:2008pw,Raidal:2016wop,Salvio:2019wcp}.  In the frequentist approach~\cite{Farhi:1989pm,Strumia:2017dvt} the last step is required for consistency~\cite{Salvio:2015gsi,Strumia:2017dvt,Salvio:2019wcp}. As a result, no negative norms appear in the theory and the S matrix is unitary~\cite{Donoghue:2019fcb}.

An alternative approach to formulate quadratic gravity as a unitary theory has been proposed in Ref.~\cite{Anselmi:2017ygm}, where different quantizations for the massless graviton and the massive spin-2 field are performed. In Ref.~\cite{Anselmi:2017ygm} the theory is formulated perturbatively in the weak-field expansion (when the metric is close to the flat metric) and in the Euclidean spacetime. The theory is then non-analytically continued to the Minkowski spacetime. The classical limit of the theory of Ref.~\cite{Anselmi:2017ygm} has been discussed in the subsequent articles~\cite{Anselmi:2018bra,Anselmi:2019rxg}. 
Given that the different quantizations for the massless and massive gravitons become the same for static configurations and when the linearization mechanism is at work the theory is perturbative, the results found in this paper regarding static metrics also apply to the unitarization procedure of Refs.~\cite{Anselmi:2017ygm,Anselmi:2018bra,Anselmi:2019rxg} discussed above\footnote{We thank D.~Anselmi for a private discussion on this subject.}. However, here we will not consider the quantum subtleties and focus on solutions within the classical theory.

Yet another approach to formulate a consistent higher-derivative theory of gravity is the one based on non-local ghost-free actions which can render the gravitational sector super-renormal- izable~\cite{Tomboulis:1997gg,Modesto:2011kw,Biswas:2011ar}. However, the UV completion of these models, which must include the absence of Landau poles, in the presence of a realistic matter sector has not (yet) been established.\footnote{See e.g. Ref.~\cite{Buoninfante:2018mre} for a recent discussion on quantum effects in ghost-free infinite derivative gravity.} That non-local gravity can result in singularity free spacetimes for point sources was first suggested in the context of string theory~\cite{Harms:1993bt,Tseytlin:1995uq}. Non-singular horizonless UCOs in non-local higher-derivative gravity have received considerable attention in the recent years~\cite{Modesto:2014eta,Li:2015bqa,Buoninfante:2018xif,Buoninfante:2018xiw,Frolov:2015bta,Koshelev:2017bxd,Buoninfante:2018rlq,Giacchini:2018wlf,Buoninfante:2019swn}. In particular, it has been observed that UCOs can be constructed in the linear regime~\cite{Frolov:2015bta,Koshelev:2017bxd,Buoninfante:2018rlq,Giacchini:2018wlf,Buoninfante:2019swn} indicating that the linearization mechanism is a general phenomenon resulting from the softening of gravity at short distances. In this context, quadratic gravity provides a simple realization of higher-derivative gravity in which these general effects may be explicitly studied.


The outline of the paper is as follows. In Sec.~\ref{SSSC} we provide a general discussion of static spherically-symmetric configurations. The linearization mechanism is then explained and proved in Sec.~\ref{Linearization mechanism}, which also presents the linear solutions. Sec.~\ref{Non-linear tests} discusses more general perfect fluid stars and compares the linear-UCO solutions previously found in Sec.~\ref{Linearization mechanism} to numerical solutions within the full non-linear theory. The stability of the UCOs is discussed in Sec.~\ref{Stability}, which addresses the potential runaway solutions due to the Ostrogradsky theorem and the relations with the metastability of the EW vacuum. The possibility of UCO DM and various related phenomenological  aspects are discussed in Sec.~\ref{sec:pheno}. We offer our conclusions in Sec.~\ref{sec:end}.

\section{Static spherically-symmetric configurations}
\label{SSSC}

Quadratic gravity features the action
\be\label{eq:HD_S}
	\mathcal{S} 
	=  \frac{1}{2\kappa}\int \td^4 x \sqrt{-g}\left(R+\frac{R^2}{6M_{0}^2}  - \frac{W^2}{2M_{2}^2} \right)
	+ \mathcal{S} _{\rm M},
\ee
where $S_{\rm M}$ denotes the matter action and $W^2\equiv W_{\mu\nu\rho\sigma}W^{\mu\nu\rho\sigma}$, where $W_{\mu\nu\rho\sigma}$ is the Weyl tensor. The positive constant $\kappa$ is related to the Newton constant $G_N$ through $\kappa \equiv 8\pi G_N$. The parameters $M_0$ and $M_2$ correspond to the masses of the extra spin-0 and spin-2 particles, respectively. Note that the coefficients of the quadratic terms are dimensionless and one can define the dimensionless combinations
\be 
	f_0\equiv \sqrt{2\kappa} M_0 , \qquad  
	f_2\equiv  \sqrt{2\kappa}M_2. 
\ee 
The gravitational field equations are
\be\label{eq:HD_eom}
	{\cal G}_{\mu\nu}\equiv G_{\mu\nu} - \frac{2}{M_{2}^2} B_{\mu\nu} + \frac{1}{3M_{0}^2} \left[R\left(R_{\mu\nu} - \frac{1}{4}R g_{\mu\nu}\right) + g_{\mu\nu} R^{;\rho}{}_{\rho} - R_{;\mu\nu} \right] = \kappa T_{\mu\nu}.
\ee
where $G_{\mu\nu}\equiv R_{\mu\nu}-g_{\mu\nu}R/2$ is the Einstein tensor, $B_{\mu\nu} \equiv \left( \nabla^{\rho} \nabla^{\sigma} + \frac{1}{2} R^{\rho\sigma} \right)W_{\mu\rho\nu\sigma}$ is the Bach tensor  and $T_{\mu\nu}$ is the (matter) energy-momentum tensor\footnote{As usual the semicolon corresponds to the covariant derivative (e.g. $R_{;\mu\nu} \equiv \nabla_\mu\nabla_\nu R$), a comma corresponds to the ordinary derivative.}.

In this work we consider the general static spherically symmetric configurations, so the general line element in Schwarzschild coordinates has the form (here we adopt the mostly plus signature for the metric)
\be\label{eq:g_gen}
	\td s^2 = - a(r)\, \td t^2  + \frac{\td r^2}{b(r)}  + r^2 (\td\theta^2 +\sin^2\theta \td\phi^2),
\ee
where $a$ and $b$ are, so far undetermined, functions of the radial coordinate $r$. The field equations for the metric components $a$ and $b$ can be obtained by inserting~\eqref{eq:g_gen} into the field equations~\eqref{eq:HD_eom} or, equivalently, by first substituting the ansatz \eqref{eq:g_gen} into the action \eqref{eq:HD_eom} and then performing the variation with respect to $a$ and $b$~\cite{Lu:2015psa}.  In all, there are two independent gravitational equations, which can be taken to be the $tt$ and the $rr$ component of~\eqref{eq:HD_eom}, while the $\theta\theta$ and $\phi\phi$ equations are satisfied due to the Bianchi identities, ${\cal G}^{\mu}{}_{\nu;\mu} = 0$. The action contains second derivatives of $a$, but only first derivatives of $b$, which already suggests that the phase space of this problem is 6-dimensional. It also follows that the highest derivatives in $tt$ equation are\footnote{ $f^{(n)}$ denotes the $n$'th derivative of $f$ with respect to $r$.} $a^{(4)}$ and $b^{(3)}$ while in the $rr$ equations they are $a^{(3)}$ and $b^{(2)}$ (see also Ref.~\cite{Lu:2015psa}). It is thus possible to use the derivative of the $rr$ equations to eliminate $a^{(4)}$ and to recast the field equations as 3-rd order differential equations of $a$ and $b$ or, after eliminating $b$, as a 6-th order differential equation for $a$ (for additional details see appendix~\ref{reduction}). This implies that 6 initial conditions are needed to determine the metric, e.g. $a$ and $b$ and their first two derivatives at the origin.

The general energy-momentum tensor compatible with staticity and spherical symmetry is given by~\cite{Mak:2001eb}
\be
	T^{\mu}_{~\nu} = {\rm diag}(-\rho,P,P_{\perp},P_{\perp})\label{GenTSS},
\ee
where the energy density $\rho$, the radial pressure $P$ and the tangential pressure $P_{\perp}$ are functions of $r$ only. The energy-momentum conservation $T^{\mu}{}_{\nu;\mu} = 0$ reads
\be\label{eq:cont}
	P' + \frac{a'}{2a}(\rho + P) + \frac{2\Delta}{r} = 0,
\ee
where we defined the pressure anisotropy parameter $\Delta \equiv P - P_{\perp} $ and a prime denotes the derivative with respect to $r$.

On top of the two gravitational field equations and the continuity equation, $P$ and $\Delta$ must be specified by two additional equations in order to close the system, or, if matter is described by classical fields, the energy-momentum tensor is determined by their action and the matter field equations imply the continuity equation \eqref{eq:cont}. For ideal fluids $\Delta = 0$ and pressure is determined from an equation of state $P = P(\rho)$. When solving the field equations for a spherical object we specify the central pressure (or density) and require that pressure vanishes at the surface of the object, located at $r={\cal R}$. This provides a physical boundary condition. We remark that the continuity equation~\eqref{eq:cont} can be integrated for ideal fluids, yielding
\be\label{GenPprof}
	\int^{P(r)}_{0} \frac{\td \bar P}{\rho(\bar P) + \bar P} = \frac{1}{2}\ln \left(\frac{a({\cal R})}{a(r)}\right).
\ee
The stress-energy tensor is taken to vanish outside of the star, i.e. there is no cosmological constant, and thus the spacetime must be asymptotically flat for physical configurations.

In this work we are interested in regular solutions sourced by matter. Indeed, a real-world curved metric should arise from a physical matter distribution. The analysis of Frobenius series of the metric functions around the origin,
\be
	a(r) = a_{t}  (r^{t} + a_{t+1} r^{t+1}+ a_{t+2} r^{t+2} + ... ), \qquad
	b(r) = b_{s}  r^{s} + b_{s+1} r^{s+1}+ b_{s+2} r^{s+2} + ...\, . 
\ee
indicates that there are three families of solutions determined by the pair $(s,t)$.\footnote{More exotic spherically symmetric solutions that may not be captured by this classification, such as wormholes, have been found numerically~\cite{Lu:2015psa} and can be found analytically in the high energy limit in which the theory becomes conformal~\cite{Hohmann:2018shl}. We will not consider such solutions here.} These are (-1,1), (0,0) and (2,2)~\cite{Stelle:1977ry,Lu:2015psa}. As solutions regular around the origin belong to the (0,0) family, all solutions considered in this paper will belong to this family. Furthermore, regularity of curvature invariants constructed from the Riemann tensor and its derivatives requires\footnote{
This can be made explicit using the coordinates $x_1= r\sin\theta\cos\varphi$, $x_2= r\sin\theta\sin\varphi$, $x_3= r\cos\theta$
\be  \label{EuMet}
	\td s^2 = -a \td t^2  +\td x_i \td x_i + \left(b^{-1}-1\right) r^{-2} x_i x_j \td x_i \td x_j.
\ee
Thus, since $a$ and $b$ are functions of $r=\sqrt{\sum_i x_i^2}$, derivatives of this metric at the origin are not well-defined unless $a$ and $b$ are even functions of  $r$ and $b(0)=1$.
}
\be\label{abrExp}
	a(r) = a_0 (1 + a_2 r^2 + a_4 r^4 + ...), \qquad
	b(r) = 1+ b_2 r^2 + b_4 r^4 ...\, , 
\ee 
with $a_0\neq 0$ and all higher-order odd terms vanishing. The regularity of the field equations around the origin, in particular of the Bach tensor, demands that the first 2 odd terms vanish. Of course, we must also require regularity of $T^{\mu}_{~\nu}$ in order for the equations to be satisfied. Then, since $a_0$ can be fixed by a suitable rescaling of the time coordinate and all higher-order coefficients can, in principle, be evaluated from the field equations, the solution depends on 2 physical free parameters $a_2$, $b_2$. As shown in the next section, in the linear regime, due to the presence of two growing modes, both parameters are fixed by asymptotic flatness. For asymptotically flat spacetimes, this must also hold in the non-linear regime as, in this case, the linear approximation must always hold in the limit $r\to \infty$. So, the absence of growing modes implies two distinct conditions. We conclude that the regular static spherically symmetric solution is unique for a given matter content of the star, which, in case of ideal fluids, can be determined by the equation of state $\rho(P)$ and the central density or pressure.

\section{Linearization mechanism}
\label{Linearization mechanism}

What we mean by ``linearization mechanism" is the following property: the spacetime is approximately flat one whenever the Schwarzschild radius
$r_h\equiv 2 G_N M$ satisfies
\be \label{BHcond}
	r_h \ll \min\left(1/M_0, 1/M_2\right)
\ee
and thus the solutions are well described by the linear perturbation theory in $h_{\mu\nu} \equiv g_{\mu\nu} - \eta_{\mu\nu}$. Importantly, when this condition is satisfied, a horizon cannot form in quadratic gravity.  This condition underlines the interplay between the Einstein-Hilbert term which, when dominant, would produce a strong gravitational field at the distance scale $r_h$, and the quadratic terms that soften gravity at length scales smaller than $1/M_0$ and $1/M_2$.

To better understand the origin of the linearization mechanism in simple terms consider first a point mass for which $T^{\mu}_{~\nu} = {\rm diag}(-M\delta(r),0,0,0)$. It generates a Newtonian potential~\cite{Stelle:1977ry}
\be\label{VNexpr}
	V_N(r) = -\frac{r_h}{2r}\left(1-\frac43 e^{-M_2 r}+\frac13 e^{-M_0 r}\right).
\ee
We can easily show that this potential satisfies\footnote{This can be shown as follows
\be
	|V_N(r)| 
	= \frac{r_h}{2r} \left|\frac43(1- e^{-M_2 r})-\frac13( 1- e^{-M_0 r})\right|
	\leq \frac{4r_h}{6r}\left| 1 - e^{-M_2 r}\right| + \frac{r_h}{6r}\left| 1 -  e^{-M_0 r}\right|
	\leq \frac{r_h}{6}(4 M_2 + M_0),
\ee
where in the last step we used $0 \leq 1- e^{-x} \leq x$.
}
\be\label{eq:Vbound_0}
	|V_N(r)| 
	\leq \frac{r_h}{6}(4 M_2 + M_0),
\ee
thus $|V_N(r)|$ is much smaller than 1 everywhere if the condition~\eqref{BHcond} is satisfied implying that the metric is approximately flat. This has been shown to be a general feature of higher-derivative linearized local gravitational theories in case the poles of the propagators are real and simple~\cite{Modesto:2014eta}.

Although the case of a point-like static source provides a simple way to understand the origin of the linearization mechanism it does not prove the mechanism. The reason is that the spacetime generated by this source is actually singular at $r=0$ and this prevents us from trusting perturbation theory in $h_{\mu\nu}$. In order to see the singularity at the origin one can simply expand the metric of the linearized theory~\cite{Stelle:1977ry},
\bea\label{aPoint} 
	a(r) &= 1- \frac{M}{4\pi r \bp^2} \left[1-\frac43 e^{-M_2 r}+\frac13 e^{-M_0 r}\right]
\\  	
	b(r) &= 1- \frac{M}{4\pi r \bp^2} \left[1-\frac23 (1+M_2 r)e^{-M_2 r}-\frac13(1+M_0 r) e^{-M_0 r}\right],
\eea
where $\bp$ is the reduced Planck mass, around $r=0$ and compare the result with the regularity condition in~\eqref{abrExp}: the absence of the odd powers of $r$ in the small $r$ expansion. Although both $a$ and $b$ in~\eqref{aPoint} are regular at $r=0$ they do not satisfy~\eqref{abrExp} and, therefore, the spacetime is {\it not} regular at $r=0$. This can be checked directly by, e.g., computing the Ricci scalar $R$ corresponding to~\eqref{aPoint} and noting that it diverges in the limit $r\to 0$ (see e.g.~\cite{Giacchini:2018wlf}).

It is important to note that this problem is not a sign of sickness of the theory but appears because a point mass is a singular source at the origin\footnote{The spacetime singularity due to a delta-function source can be avoided in non-local higher-derivative gravity (see e.g.~\cite{Biswas:2011ar}).}. Indeed, the same problem emerges in an asymptotically free, and thus UV-complete, Yang-Mills theory with a $\delta$-function current: for instance, taking this current to flow along the time direction in the spacetime and along a fixed direction in the Lie algebra of the Yang-Mills gauge group. With this choice there always exists a singular Abelian solution along the chosen Lie-algebra direction: this happens for the same technical reason why a $\delta$-function charge distribution generates the singular $1/r$ potential in classical electrodynamics. Therefore, in order to fix this issue regular sources must be considered.

To understand the linearization mechanism in more general terms consider  the field equations~\eqref{eq:HD_eom} at the leading order in the metric perturbation $h_{\mu\nu}$ , which can be expressed as
\be\label{LinEq}
	K^{\mu\nu\rho\sigma} h_{\rho\sigma} = \kappa T^{\mu\nu},
\ee
where $K^{\mu\nu\rho\sigma}$ is a differential operator that includes derivatives up to the fourth order (see e.g.~\cite{Salvio:2014soa}).  Due to diffeomorphism invariance of the full equations, the linear equations in~(\ref{LinEq}) are invariant under the gauge transformation $h_{\mu\nu} \to h_{\mu\nu} + \xi_{\mu,\nu} + \xi_{\nu,\mu}$, where $ \xi_{\mu}$ are generic functions of the spacetime point.

As usual when expanding around the flat Minkowski spacetime one can find a solution in terms of a Fourier transform. For simplicity consider stationary configurations. Requiring asymptotic flatness, the momentum space metric induced by matter is given by
\be 
	\tilde h_{\mu\nu}(k) = \kappa iD_{\mu\nu\rho\sigma}(k) {\tilde T}^{\rho\sigma}(k),  
\ee
where $D_{\mu\nu\rho\sigma}(k)$ represents the momentum-space propagator for $h_{\mu\nu}$, the four-momentum $k$ is here given by $k= (0, \vec k)$ as we are dealing with static configurations, and
$\tilde X(\vec k) \equiv \int \td^3 x \, \exp(-i \vec k\cdot  \vec x) X(\vec x)$ is the spatial Fourier transform of a generic field $F(\vec x)$.  Inserting the well-known explicit expression of $D_{\mu\nu\rho\sigma}(k)$ (see e.g.~\cite{Salvio:2014soa}) into the above equation then gives
\be\label{eq:h_sol_D}
	\tilde h_{\mu\nu}(k)
	=  \frac{1}{3} \frac{\kappa M_{0}^2 \tilde T \eta_{\mu\nu} }{k^2 (k^2 + M_{0}^2)} - \frac{2\kappa M_{2}^2\left( \tilde T_{\mu\nu} - \frac{1}{3} \tilde T \eta_{\mu\nu}  \right)}{k^2 (k^2 + M_{2}^2)}  + \left(\ldots \right) k_{\mu} k_{\nu}, 
\ee
where $T \equiv T^{\mu}_{~\mu}$ and the ellipsis stands for terms that can be set to zero by a suitable gauge choice. Now, for a generic field $X(\vec x)$ we have
\bea
	\left| \int \frac{\td^3 k}{(2\pi)^3} e^{i \vec k \vec x}  \frac{\tilde X(k)  m^2}{k^2 (k^2 + m^2)} \right|	
\leq	 |X|_{\rm int}  \int \frac{\td^3 k}{(2\pi)^3} \frac{m^2}{k^2 (k^2 + m^2)}  
=	  \frac{m |X|_{\rm int} } {16 \pi^2}
\eea
when $ |X|_{\rm int} < \infty$. To find this result we used $|\tilde X(\vec k)| \leq |X|_{\rm int}$ and defined
\be
	|X|_{\rm int} \equiv \int \td^3 x |X(\vec x)|.
\ee
It follows that 
\be\label{eq:h_bound}
	|h^{\mu}{}_{\nu}| \leq \frac{\kappa }{48 \pi^2} M_{0} |T|_{\rm int} \delta^{\mu}_{\nu}+ \frac{\kappa }{8 \pi^2} M_{2} \left| T^{\mu}{}_{\nu} -  T\delta^{\mu}_{\nu}/3 \right|_{\rm int} + \left(\ldots \right).
\ee
So, up to gauge dependent terms denoted by the ellipsis (which can be set to zero), the metric perturbation is bounded for finite non-singular sources. If we further assume that $\rho > 0$, and the components of stress energy tensor are bounded by $\rho$ , i.e. there exists a constant $C>0$ so that $|T_{\mu\nu}| \leq C \rho$, e.g. when the dominant energy condition is satisfied, then an upper bound analogous to \eqref{eq:Vbound_0} holds and thus the condition in \eqref{BHcond} implies the weak-field limit.

To provide more concrete examples, we will now turn our attention to spherically symmetric sources. Spherical symmetry together with energy-momentum conservation, which at the leading order reads $0 = k_i \tilde{T}^{ij}$, further implies the following tensorial decomposition\footnote{Direct evaluation of the Fourier transform gives
\bea
	\tilde T_{\mu\nu}(\vec k)
	&= -\tilde \rho  \, \frac{p_{\mu} p_{\nu}}{p^2} 
	+ \left(\tilde P - \frac{\tilde \Delta + \tilde \Delta_r} {2}\right) \left( \eta_{\mu\nu} -  \frac{p_{\mu} p_{\nu}}{p^2} \right) - \frac{\tilde \Delta - 3\tilde \Delta_r} {2} \frac{k_{\mu} k_{\nu}}{k^2},
\eea
where 
$
	\tilde\Delta_r(|\vec k|) \equiv \int \td^3 x \, e^{-i \vec k\cdot  \vec x} (\hat k \hat x)^2 \Delta(r).
$
Eq.~\eqref{eq:tildeT} is obtained by using $0 = k_i \tilde{T}^{ij} = \tilde P - \tilde\Delta+\tilde\Delta_{r}$.}
\bea\label{eq:tildeT}
	\tilde T_{\mu\nu}(\vec k)
	&= -\tilde \rho  \, \frac{p_{\mu} p_{\nu}}{p^2}  
	+ \frac{1}{2}\left( \eta_{\mu\nu} -  \frac{p_{\mu} p_{\nu}}{p^2}  - \frac{k_{\mu} k_{\nu}}{k^2} \right)  \tilde T^{i}{}_{i},
\eea
where $p_{\mu} = \delta_{\mu}^{0}$.
Note that the continuity equation~\eqref{eq:cont} implies $\tilde T^{i}{}_{i} \equiv 3\tilde P -2 \tilde \Delta = -k \tilde{P}_{,k}$, with $k\equiv |\vec k|$.

Let us look now at the shape of $h_{\mu\nu}$, which is a gauge dependent quantity. Due to spherical symmetry, in coordinate space \eqref{eq:h_sol_D} has the general form $h_{ij}(\vec x) = A(r) \delta_{ij} + B(r) \hat x_i \hat x_j$, where $\hat x_i$ is the radial unit vector and $A$ and $B$ are some functions of $r$. However, linearising the spherically symmetric metric \eqref{eq:g_gen}, 
\be
	\delta a \equiv a - 1 = -h_{tt}, \qquad  \delta b \equiv b - 1 \approx - h_{rr},
\ee
implies that $h_{ij}(\vec x) \propto \hat x_i \hat x_j$. Thus, in order to use a coordinate system compatible with the one  in \eqref{eq:g_gen} we choose the gauge dependent piece in \eqref{eq:h_sol_D} so that $A = 0$. In momentum space, the spatial metric tensor has the general form $\tilde h_{ij}(\vec k) = \tilde{\mathcal{A}}(|\vec k|)\delta_{ij} + \tilde{\mathcal{B}}(|\vec k|) k_i k_j$, so in configuration space $h_{ij}(\vec x) = \mathcal{A}(r)\delta_{ij} - \mathcal{B}(r)_{,ij}$, where $\mathcal{A}$ and $\mathcal{B}$ are some functions of $r$, not necessarily equal to $A$ and $B$. Using gauge freedom we can choose $\mathcal{B}(r)$ so that $h_{ij}(\vec x) \propto \hat x_i \hat x_j$, that is, $A = 0$, which implies that $h_{ij}(\vec x) = -\hat x_i \hat x_j r \partial_r \mathcal{A}(r)$ in this gauge. 

With this choice we obtain 
\bea
	\delta a 
	&= -\kappa \int \frac{\td^3 k}{(2\pi)^3} e^{i \vec k \vec x} 
	\left( \frac{\tilde \rho + \tilde T^{i}{}_{i}}{k^2} + \frac{\frac{1}{3}\tilde \rho - \frac{1}{3}\tilde T^{i}{}_{i} }{k^2 + M_{0}^2} + \frac{-\frac{4}{3} \tilde\rho - \frac{2}{3} \tilde T^{i}{}_{i}}{k^2 + M_{2}^2}  \right),  
	\\
	\delta b 
	&=  \kappa  r \partial_r \int \frac{\td^3 k}{(2\pi)^3} e^{i \vec k \vec x} 
	\left(  \frac{\tilde \rho}{k^2} + \frac{-\frac{1}{3}\tilde\rho+\dfrac{1}{3} \tilde T^{i}{}_{i}}{k^2 + M_{0}^2} + \frac{-\frac{2}{3} \tilde\rho-\frac{1}{3} \tilde T^{i}{}_{i}}{k^2 + M_{2}^2} \right),
\eea
so the metric perturbation can be expressed through a combination of Yukawa like potentials
\bea\label{eq:da_db}
	\delta a/2 &= V(r;0)  + \frac{1}{3}V(r;M_0) - \frac{4}{3}V(r;M_2) + U(r;0) - \frac{1}{3}U(r;M_0)  - \frac{2}{3} U(r;M_2) ,
	\\ 
	\delta b/2 &= -r V'(r;0) + \frac{1}{3} rV'(r;M_0) + \frac{2}{3} rV'(r;M_2) - \frac{1}{3} r U'(r;M_0) + \frac{1}{3} r U'(r;M_2), 
\eea
where we defined 
\bea\label{eq:VU}
	 V(r;m) 
&	\equiv -\frac{\kappa }{2} \int \frac{\td^3 k}{(2\pi)^3} \frac{\tilde \rho e^{i \vec k \vec r} }{k^2 + m^2} 
	 = -G_{N} \int   \td^3 x \frac{e^{- m |\vec x-\vec r|}}{|\vec x-\vec r|} \rho(\vec x) , 
\\	
	 U(r;m) 
&	\equiv -\frac{\kappa }{2} \int \frac{\td^3 k}{(2\pi)^3} \frac{\tilde T^{i}{}_{i} e^{i \vec k \vec r} }{k^2 + m^2}  
	 = - G_{N} \int  \td^3 x \frac{e^{- m |\vec x-\vec r|}}{|\vec x-\vec r|} \nabla_{\vec x} \cdot \left(\vec x P(\vec x)\right).
\eea
Note that the pressure anisotropy $\Delta$ does not appear because it was eliminated using the leading order continuity equation $P' + 2 \Delta/r=0$.  Eqs.~\eqref{eq:da_db} and~\eqref{eq:VU} provide (for the first time to the best of our knowledge) the asymptotically flat solution sourced by a general static and spherically symmetric energy-momentum tensor.

Some examples of static spherically symmetric configurations: 
\begin{itemize}
	\item	The point-like source $\rho = M \delta(\vec x)$, $P=\Delta=0$ mentioned above corresponds to
\be
	V_{\rm point}(r;m) = -\frac{G_{N} M}{r} e^{- m r}
\ee
and $U = 0$ due to vanishing pressure.

	\item	 A  spherical shell of radius ${\cal R}$, mass $M$, that is $\rho = \frac{M}{4\pi  {\cal R}^2} \delta(r -  {\cal R})$ and pressure $P = P_i \delta(r -  {\cal R})$, corresponds to
\bea\label{eq:V_shell}
	V_{\rm shell}(r;m) &= 
	-G_{N} M \left\{\begin{array}{ll}
	\frac{1}{{\cal R}} e^{-m{\cal R}}\frac{\sinh (r m)}{r m},  & r < {\cal R}	\\
	\frac{1}{r} e^{-mr}\frac{\sinh ({\cal R} m)}{{\cal R} m},	& r \geq {\cal R}
	\end{array}\right. 
	\\
	U_{\rm shell}(r;m) &= 
	-4\pi G_{N} P_i  {\cal R}^2 \left\{\begin{array}{ll}
	\frac{1 + m {\cal R}}{{\cal R}} e^{-m{\cal R}}\frac{\sinh (r m)}{r m}, 				& r < {\cal R}	\\
	\frac{1}{r} e^{-mr} \frac{\sinh ({\cal R} m) - m {\cal R}\cosh ({\cal R} m)}{m {\cal R}},	& r \geq {\cal R}
	\end{array}\right. 
\eea
The continuity equation implies anisotropic pressure at the shell. Although it is a highly idealized matter configuration, it shows how each shell of a general spherically symmetric matter distribution contributes to the linear potential.

\item A sphere of radius ${\cal R}$, mass $M$ and constant internal density $\rho_i$ and pressure $P_i$ corresponds to
\bea\label{eq:V_ball}
	V_{\rm ball}(r;m) &=
	- G_{N} M 
	\left\{\begin{array}{ll}
	\frac{3}{{\cal R}^3 m^2} \left(1-\frac{\sinh (r m)}{r m} e^{-{\cal R} m} (1 + {\cal R} m ) \right)	&, r < {\cal R}	\\
	\frac{3}{r} e^{-mr} \frac{ {\cal R} m \cosh({\cal R} m) - \sinh ({\cal R} m) }{({\cal R} m)^3}	&, r \geq {\cal R}
	\end{array}\right. 
	\\
	U_{\rm ball}(r;m) &=
	-4\pi G_{N} P_i {\cal R}^3\left\{\begin{array}{ll}
	\frac{3}{{\cal R}^3 m^2} \left(1-\frac{\sinh (r m)}{r m} e^{-{\cal R} m} (1 + {\cal R} m + ({\cal R} m)^2/3 ) \right)	&, r < {\cal R}	\\
	\frac{3}{r} e^{-mr} \frac{ {\cal R} m \cosh({\cal R} m) - (1+({\cal R} m)^2/3)\sinh ({\cal R} m) }{({\cal R} m)^3}	&, r \geq {\cal R}
	\end{array}\right. 
\eea
Note that the continuity equation implies a shell of anisotropic pressure at the surface, $\Delta =  P_i {\cal R} \delta(r -  {\cal R})/2$.
\end{itemize}
Alternatively, the metric can be obtained by directly solving the field equations by imposing regularity at the origin and spatial infinity and  appropriate conditions on the objects boundary. The latter approach was used in Refs.~\cite{Stelle:1977ry,Lu:2015psa} to derive the linearized metric for the shell and ball sources given above.

Eq.~\eqref{eq:da_db} gives the metric perturbation for a given distribution for matter. In case the latter is unknown it must be first determined from the corresponding equations of motions. For ideal fluids in the non-relativistic limit, i.e. when $P \ll \rho$, the continuity equation at leading order in the metric perturbation reads 
\be\label{eq:dP_weak}
	P' = -\frac{\delta a'}{2} \rho.
\ee
This is simply the equation for hydrodynamic equilibrium, that in the GR limit takes the known form $P'  = G_{N} M(r) \rho(r)/r^2$, where $M(r)$ is the mass enclosed within radius $r$. To determine the matter configuration in this case one can now solve \eqref{eq:dP_weak} together with the gravitational field equations with the boundary condition $P({\cal R}) = 0$.

If the density profile is known, then, in the non-relativistic limit case, if $P \ll \rho$, pressure can be estimated perturbatively by setting it to zero at the leading order and then integrating \eqref{eq:dP_weak} to obtain the leading contribution to pressure. At the leading order, this is equivalent to using Eq.~\eqref{eq:cont}.

Both the shell and the ball solutions above are, unlike the point-like source, regular at $r=0$ because the small $r$ expansion in~\eqref{abrExp} is satisfied in both cases. Therefore, these sources provide explicit examples of the linearization mechanism. 

For general spherically symmetric metrics the bound \eqref{eq:h_bound} takes the form (see appendix \ref{bounds})
\bea\label{abLin}
	|\delta a(r)/2| & < 2G_N |\rho|_{\rm int} \frac{M_{0}  + 4 M_{2}}{3} + 2G_N  |P|_{\rm int} \frac{M_{0}  + 2M_{2}}{9},
	\\
	|\delta b(r)/2| & < 2G_N |\rho|_{\rm int} \frac{M_{0}  + 2 M_{2}}{9} + 2G_N|P|_{\rm int} \frac{M_{0}  + M_{2}}{6}.
\eea
By assuming the dominant energy condition,  $|P|\leq \rho$, we obtain that $|P|_{\rm int} \leq |\rho|_{\rm int}  = M$ and an inequality similar to \eqref{eq:Vbound_0} follows,
\be
	|\delta a(r)/2| \leq r_h \frac{4 M_{0}  + 14 M_{2}}{9},	\qquad
	|\delta b(r)/2| \leq r_h \frac{5 M_{0}  + 7 M_{2}}{18}.
\ee
As the integrals above depend on the mass of the object, but not its size, we see that compact objects, ${\cal R}<r_h$, can generate a very weak gravitational field as long as~\eqref{BHcond} holds. So, a horizon cannot form.

Curvature invariants depend also on derivatives of the metric, which, in the linear case, will appear through derivatives of the potentials  given in \eqref{eq:VU}. Having the linear metric for the constant density star at hand, we can directly compute the curvature invariants for the metric given by $a = \exp(\delta a)$, $b = 1 + \delta b$.\footnote{There are several ways to extend the linear solution beyond the leading order. With this choice, the gravitational field equations are given by polynomials of $\delta a$ and $\delta b$ and their derivatives. } For example, we obtain that, the Ricci scalar grows as $R \sim (r_h/{\cal R}) \times \mathcal{O}(M_{0,2}^4)$ while the square of the Weyl tensor behaves as $W^2 \sim (r_h/{\cal R})^2 (r/r_h)^4 \times \mathcal{O}(M_{0,2}^2)$ in the limit when ${\cal R} \to 0$ with $r_h$ kept constant. However, the leading order deviation from the exact field equations \eqref{eq:HD_eom} around the origin is $\mathcal{G}^{t}{}_{t} \sim \rho [1+ r_h {\cal R}\times\mathcal{O}(M_{0,2}^2) ]$ and $\mathcal{G}^{r}{}_{r} \sim \rho r_h {\cal R}\times\mathcal{O}(M_{0,2}^2) $ in the limit ${\cal R} \to 0$, so the linear solution tends to be more accurate around the origin as ${\cal R}$ decreases.  

As mentioned in the introduction, in Ref.~\cite{Salvio:2017qkx} it was shown that quadratic gravity can hold up to infinite energy flowing to conformal gravity in the infinite energy limit. Since the $R^2$ term in the action breaks the Weyl invariance of conformal gravity this means  $\sqrt{2}M_0/\bp = f_0\to \infty$ in the UV. Then, one might worry that, if this is realized,  $r_h M_0$ could go to infinity and then~(\ref{BHcond})  would never be satisfied. However, this does not happen.
This is because the $\beta$-function of $f_0$ is $\beta(f_0) = {\cal O}(1/f_0)$ at large energy $1/r_h$  as shown in~\cite{Salvio:2017qkx}. This implies that   the coupling $f_0$ grows actually very slowly as $r_h\to 0$:
\be  
	f_0\lesssim \sqrt{\ln(\bar r/r_h)}, 
\ee
where $\bar r$ is some reference length scale. So when $r_h\to 0$ 
\be 
	r_h M_0\lesssim r_h \sqrt{\ln(\bar r/r_h)}  \bp
\ee 
and~(\ref{BHcond}) can be satisfied. The linearization condition in~\eqref{BHcond} is, therefore, compatible with the theory being UV complete.

\section{Towards realistic stars in the non-linear theory}
\label{Non-linear tests}
 
In this section we perform numerical non-linear calculations for a constant density star and discuss generalizations to more realistic matter configurations. One of the motivations is explicitly testing the linearization mechanism in the full theory.
 
The field equations \eqref{eq:HD_eom} are solved numerically by imposing regularity at the origin, which enforces the expansion~\eqref{abrExp}, and asymptotic flatness.
As was described in section~\ref{SSSC}, the coefficient $a_0$ is arbitrary, while the coefficients $a_2$ and $b_2$ are fixed by the condition of asymptotic flatness: i.e. $a' \to 0$ and $b \to 1$ as $r \to \infty$; this is achieved via the shooting method.  In the numerical implementation we impose the initial conditions at a finite value of $r$ and, for greater accuracy, we approximate the solution around the origin by an expansion of the form \eqref{eq:HD_eom} with a few coefficients $a_k$ and $b_k$ with $k\geq 4$ computed from the expanded field equations.  As we integrate outwards, the surface of the object in question corresponds to the sphere at which the pressure vanishes. 

\begin{figure}[t]
\begin{center}
\includegraphics[scale=0.54]{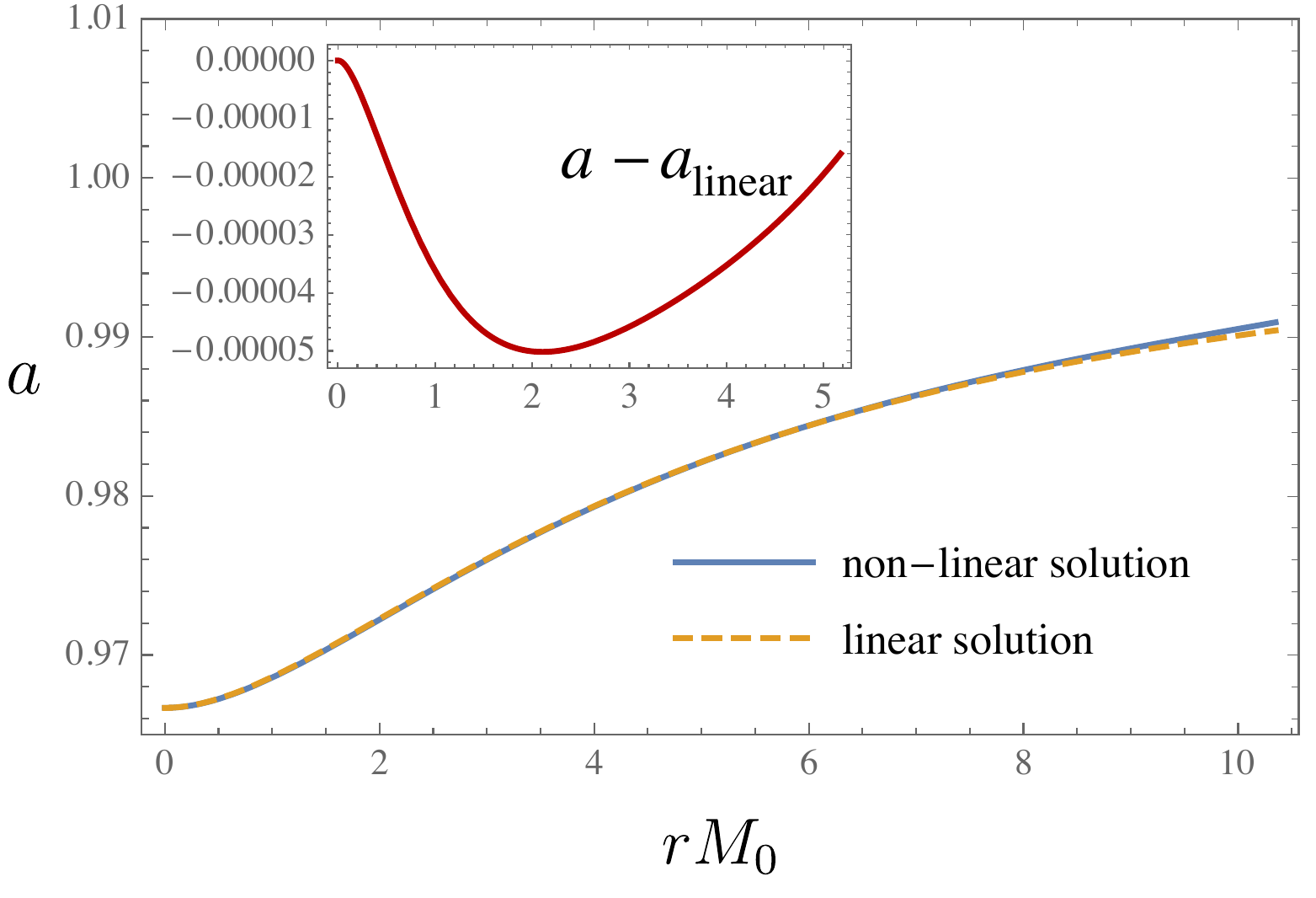}   \,    \includegraphics[scale=0.54]{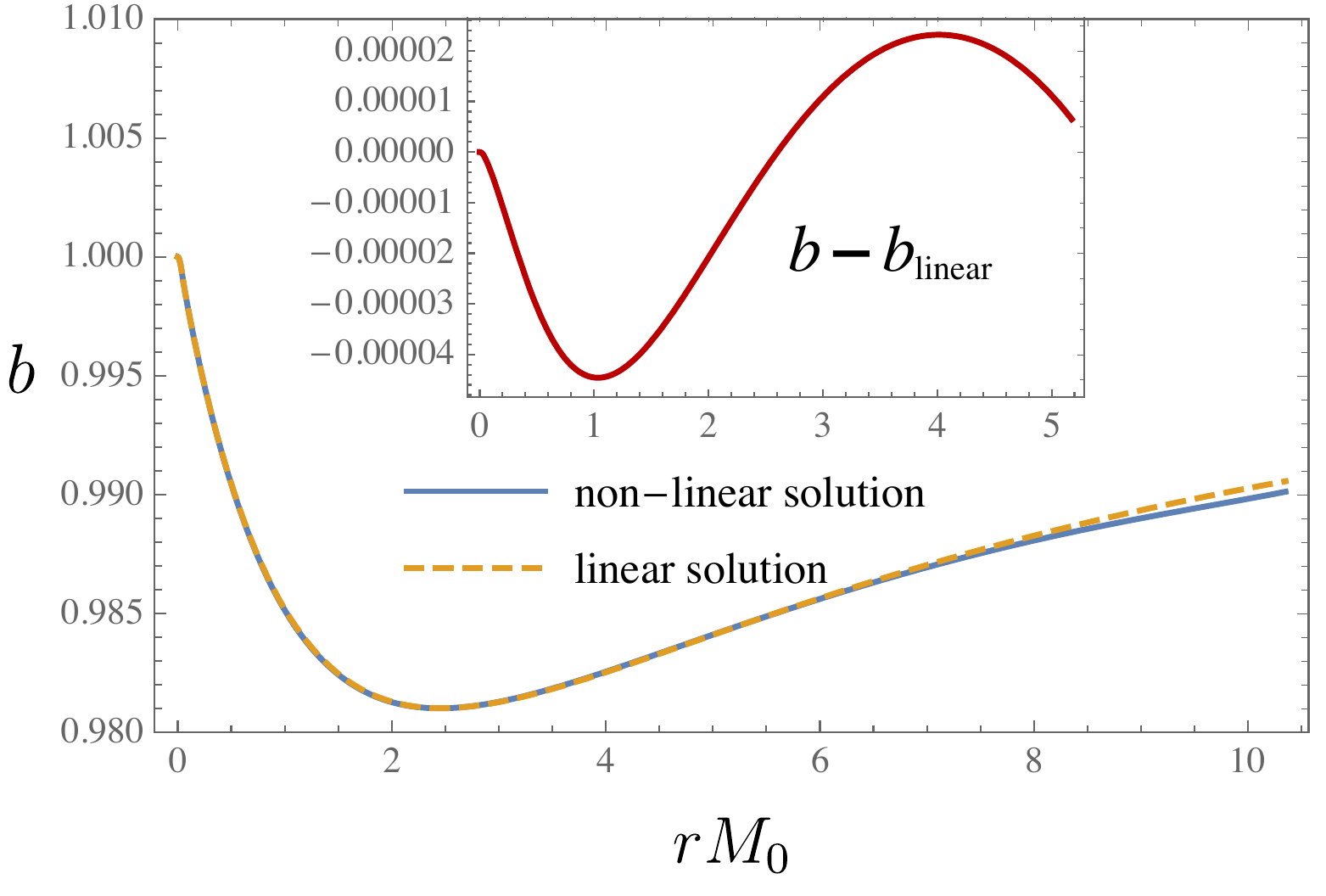}
\end{center}
	\caption{\em  The linear analytic solution  and the non-linear numerical solution when the linearization condition in~(\ref{BHcond}) holds.  The insets  show the very small difference between the linear analytic solution  and the non-linear numerical solution.
We set $M_2 = M_0/2$,   $f_2 =10^{-8}$,  the radius ${\cal R}$ of the constant density object to be $0.05/M_0$ and the Einsteinian horizon $r_h=0.1/M_0 > {\cal R}$ (so that in GR in would be a BH).}
\label{abr}
\end{figure}
\begin{figure}[t]
\begin{center}
\includegraphics[scale=0.52]{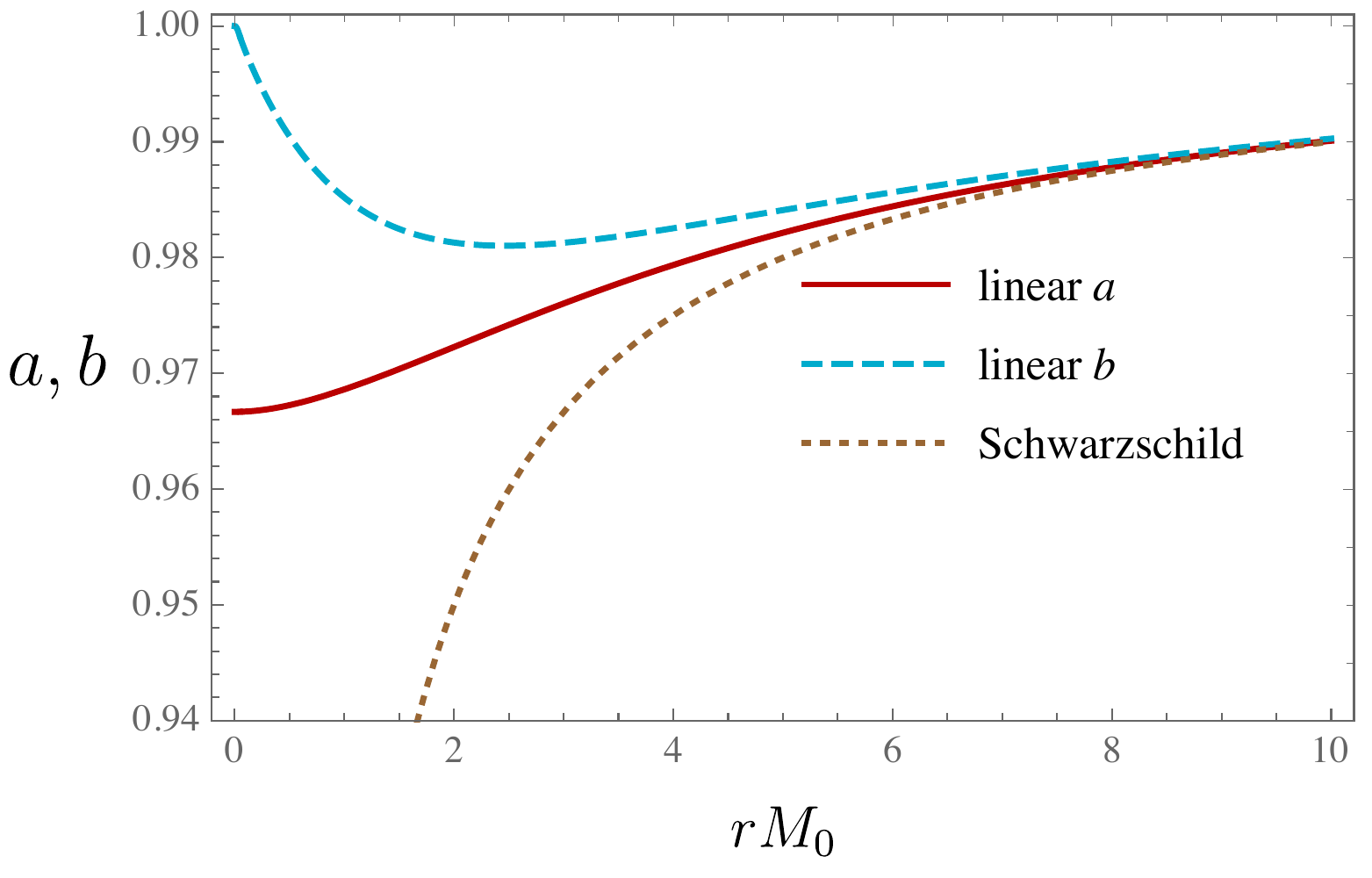}  \,    \includegraphics[scale=0.56]{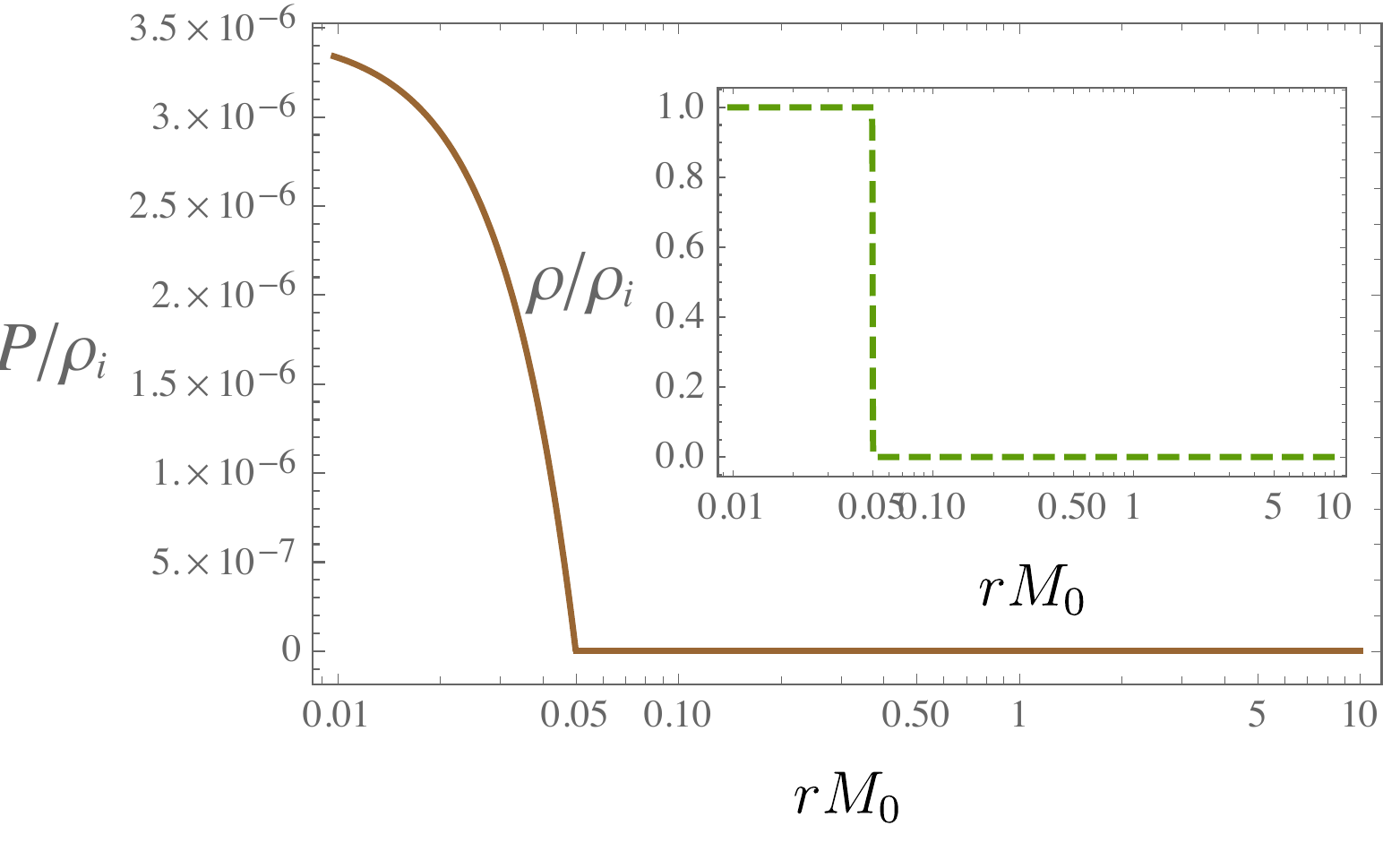}
\\ 
\vspace{1cm}
 \includegraphics[scale=0.46]{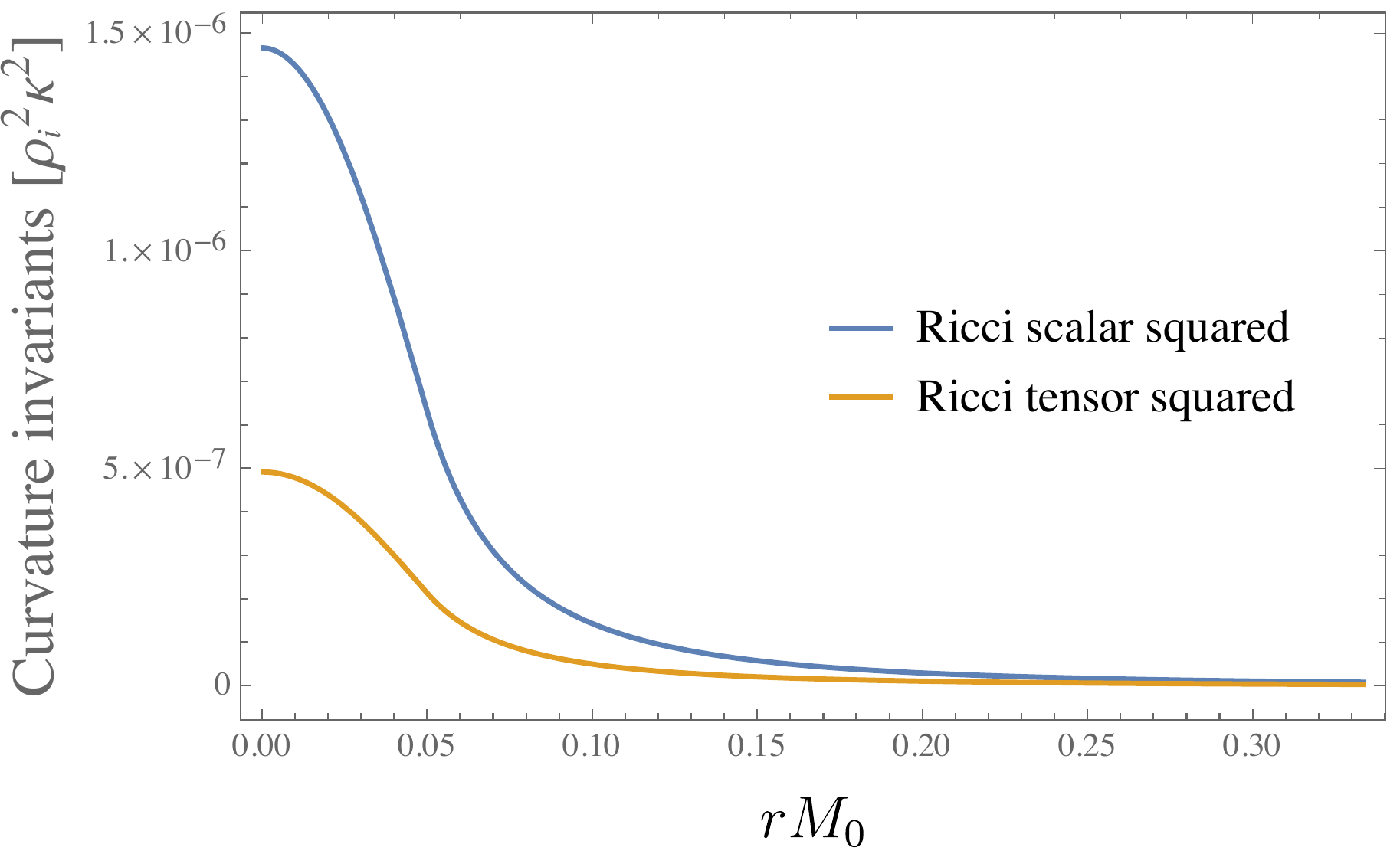}  \,\,\,\,   \includegraphics[scale=0.45]{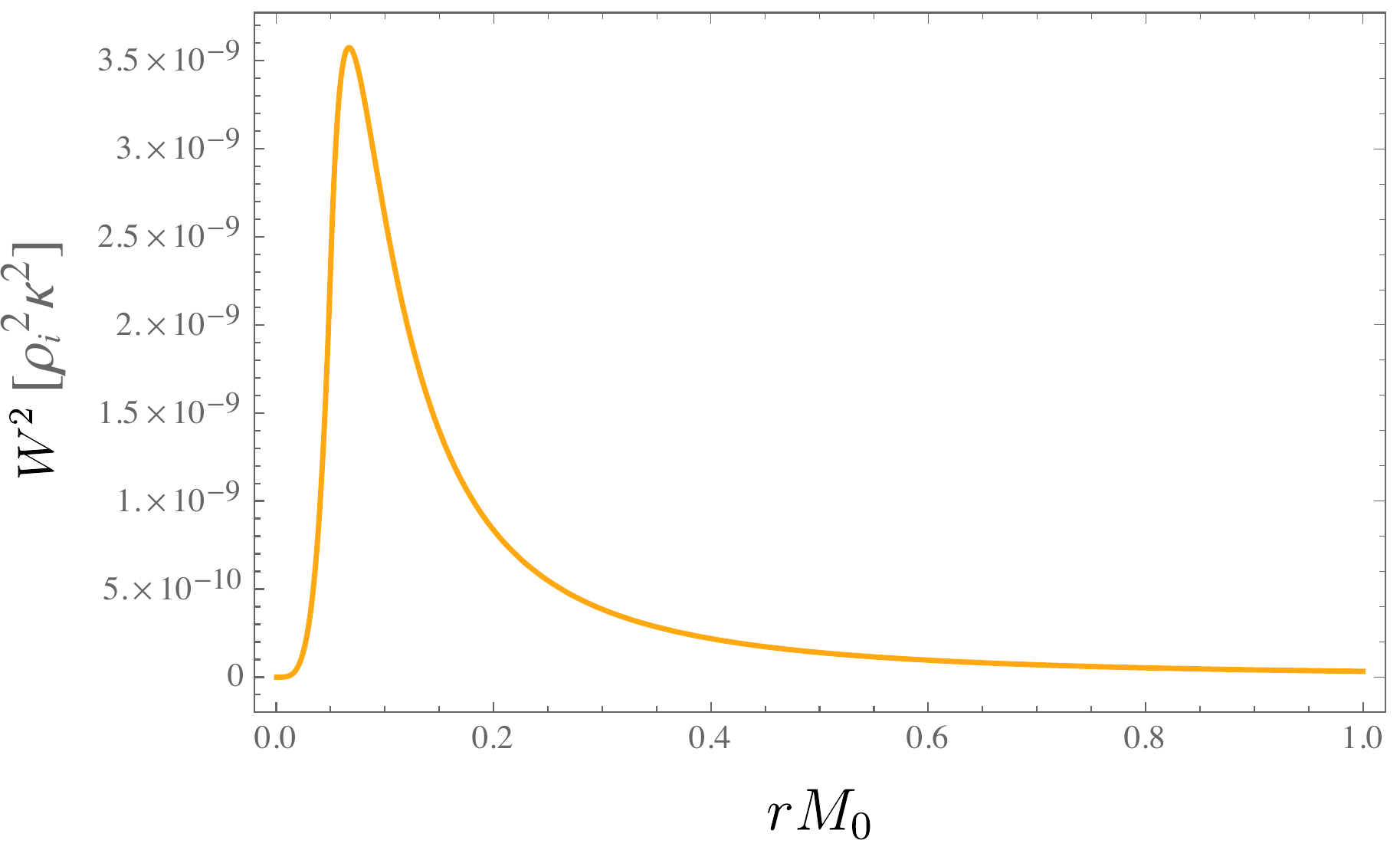}
\end{center}
	\caption{\em  Comparison with the Schwarzschild solution of GR (upper left) and the  pressure profile (upper right). The bottom plots show some curvature invariants: the Ricci scalar squared $R^2$, the Ricci tensor squared $R_{\mu\nu}R^{\mu\nu}$ and the Weyl tensor squared $W^2$.  The parameters are set as in Fig.~\ref{abr}.}
\label{abr2}
\end{figure}

An example which we treat in some detail is the constant density ball. This useful idealized toy model can be characterized by an equation of state $\rho = \rho_i$ of an incompressible fluid. As the speed of sound $c_s^2 = \td P/\td \rho$ within this star is infinite, this configuration is not completely realistic. However, in the non-relativistic case, i.e. when $P \ll \rho$, it approximates well stars comprised of matter with equation of state $\rho = \rho_i + P/c_s^2$, where $c_s^2$ is a constant (we checked that numerically). This behavior can be easily understood by noting that when $P \ll \rho_i < \rho$, the metric perturbations are determined mainly by the density and thus pressure appears as a small correction. Although $\rho = \rho_i + P/c_s^2$ satisfies causality for $c_s^2 <1$, it implies a sharp in density at the boundary and could thus be refined further. A well-motivated example of such an equation of state is provided by the bag model~\cite{Chodos:1974je} for quark gluon plasma describing superdense baryonic corresponds to $c_s^2 = 1/3$ and $\rho_i = \mathcal{O}(100 \, {\rm MeV}/{\rm fm}^3)$ ~\cite{Buballa:2003qv}. 

To compare our results with the corresponding stars in GR, we note that numerical studies indicate the maximally stiff equations of state $\rho = \rho_i + P/c_s^2$ generate the most compact stars for a given upper bound on $c_s^2$~\cite{Urbano:2018nrs}. The $c_s \to \infty$ limit, i.e. the constant density star, allows the saturation of the Buchdahl limit~\cite{Buchdahl:1959zz}
\be
	\frac{r_h}{{\cal R}}< \frac{8}{9}, 	\label{BuchL}
\ee 
satisfied in GR for all static spherically symmetric configurations of ideal fluids. When restricting to causal equations of states, with $c_s^2 \leq 1$, slightly less compact stars are allowed $r_h/{\cal R}< 0.71$~\cite{1984ApJ...278..364L,Glendenning:1992dr}.  The Buchdahl limit can be generalized to anisotropic stars with a positive pressure anisotropy $\Delta >0$ as long as the density within the star decreases~\cite{Guven:1999wm}. This includes boson stars with a monotonous potential~\cite{Urbano:2018nrs}. However, when these conditions are violated then anisotropic stars can violate the Buchdahl limit and may approach the compactness of a BH~\cite{Raposo:2018rjn}, which provides the ultimate limit on compactness in GR.

To construct the analytic solutions, we assume a vanishing pressure at the leading order, so the linear metric corresponds to~\eqref{eq:V_ball} with $U(r,m) = 0$. 
The pressure for the linear solution can be obtained perturbatively from~\eqref{GenPprof}, which for a constant density gives 
\be
	P(r) =  \rho \left( \sqrt{a({\cal R})/a(r)}- 1 \right) \label{Psol},
\ee
where for $a$ we can use the leading order linear metric. Treating $\rho$ as a formal expansion parameter we see that $P =  \mathcal{O}(\rho^2)$ because $a(r) = 1 + \mathcal{O}(\rho)$. At the non-linear level the pressure must be obtained by solving the continuity equation~\eqref{eq:cont} together with the gravitational field equations \eqref{eq:HD_eom}. We remark that non-negativity of pressure implies that $a(r) \leq a({\cal R})$. For certain combinations of masses this condition could be violated at the leading order around the origin, in which case thick shell solutions might be constructed with positive pressure inside the shell and zero pressure at the boundary of the shell. 

 In Fig.~\ref{abr} we compare the non-linear numerical solution with the linear solution when the latter should be a good approximation according to the linearization mechanism, that is when~(\ref{BHcond}) holds. We take the constant density ball for the sake of simplicity. The agreement of the two solution is remarkable as shown in the insets, which give  the very small difference between these two solutions.
In Fig.~\ref{abr2} one can see that both $a$ and $b$ reduce to the Schwarzschild solution at large distances, that is for $r\gg \max(1/M_2,1/M_0)$. The profile of the pressure $P$ computed through~(\ref{Psol}) is also provided; $P$ is positive everywhere. In Figs.~\ref{abr} and~\ref{abr2} we have chosen ${\cal R} < r_h$, which not only violates the Buchdahl limit~\eqref{BuchL} but is also more compact than a Schwarzschild BH. We see that a horizon does not form in quadratic gravity as the linear theory is a very good approximation even for objects more compact than a Schwarzschild BH.

We remark that the square of the Weyl tensor, shown in Fig.~\ref{abr}, vanishes near the origin indicating conformal flatness there. In fact, it can be shown analytically that the Weyl tensor scales as $r^2$ around the origin when the metrics admits an expansion \eqref{abrExp} as is the case for the interior linear metric. This property also exists in non-local higher-derivative gravity~\cite{Buoninfante:2018xiw}.

\section{Stability and metastability}\label{Stability}

Let us start this section discussing the stability of the UCOs found in Sec.~\ref{Linearization mechanism}. As shown in Sec.~\ref{SSSC}, for a given matter content of the star, currently determined by the equation of state $\rho(P)$ and central pressure (or density), there is a unique horizonless static spherically-symmetric solution that is asymptotically flat and regular at the origin. The constant density incompressible fluid UCOs are, by construction, stable against spherical perturbations. A stable finite size solution is also expected to exist for compressible fluids, because the gravitational pull weakens at small distances and thus a collapse to a point can be avoided even at low pressure. We expect that the size of stable objects that would collapse in GR is dynamically set at the scale where gravity is softened, that is, somewhere around or below $1/M_2$ or $1/M_0$. However, testing that requires a dedicated study of the non-linear equations, which is beyond the scope of this work.

We now analyze  the stability of these very compact solutions under non-linear Ostrogradsky runaways. First note that
\be
	\frac{r_h}{2G_N} = M = \frac{4\pi}{3} {\cal R}^3 \bar\rho,  
\ee
where $\bar\rho\equiv {\cal V}^{-1} \int d^3 x \, \rho(\vec x)$ is the average energy density, ${\cal V}$ is  the volume of the compact object in question and ${\cal R}$ is its linear size, which is defined in terms of ${\cal V}$ by ${\cal R}\equiv \sqrt[3]{3{\cal V}/(4\pi)}$. 
So, by using~\eqref{BHcond} and remembering ${\cal R}<r_h$ for very compact objects
\be  
	\bar\rho 
	= \frac{3r_h}{8\pi G_N {\cal R}^3} 
	= \frac{3r_h \bp^2}{{\cal R}^3} \gg \frac{3r_h}{{\cal R}} M_2^2 \bp^2 .
	 \label{rhoM2}
\ee 
On the other hand, in order to avoid the Ostrogradsky instabilities we need that the matter energy density be bounded by (see Eq.~(23) of~\cite{Salvio:2019ewf})
\be 
	\rho \ll M_2 \bp^3. 
\ee 
In order to be conservative we demand this condition to be fulfilled even for the maximal energy density
\be  
	\rho_m\equiv \max_{\vec x}\rho(\vec x) \ll M_2 \bp^3. 
	\label{rhombound}
\ee
Combining~(\ref{rhoM2}) and~(\ref{rhombound}) we obtain
\be 
	\frac{3r_h}{{\cal R}} M_2^2 \bp^2
	\ll \bar\rho\leq\rho_m\ll 
	M_2 \bp^3. 
\ee 
We observe that these conditions can be satisfied 
when
\be 
	M_2 \ll \frac{{\cal R}}{3r_h}\bp, \label{Crun}
\ee
which is easily satisfied for values of $M_2$ that correspond to a natural Higgs mass. Indeed, in Ref.~\cite{Salvio:2014soa,Salvio:2017qkx} it was shown that the Higgs mass $M_h$ can be naturally set to its observed value $M^{\rm exp}_h$ for $M_2\lesssim 10^{10}$~GeV. This is because the gravitational quantum correction to $M_h$ is not much larger than $M^{\rm exp}_h$ in this case. The reason why this happens is because with this setup the Higgs field acquires a shift symmetry softly broken at the scale $M^{\rm exp}_h$ or below. Note that condition~(\ref{Crun}) does not need to apply to the unitarization mechanism of Ref.~\cite{Anselmi:2017ygm} because in that case the classical theory might not have runaway solutions at all~\cite{Anselmi:2018bra,Anselmi:2019rxg}.
 
Another interesting feature of the UCOs and the linearization condition in~(\ref{BHcond}) is that it can avoid the formation of microscopic BHs with horizons $r_h$ smaller than $1/h_{\rm max}$, where $h_{\rm max}$ is the value of the Higgs field for which the effective Higgs potential acquires its maximum. Indeed, these microscopic BHs have been proven to be very dangerous for the Standard Model as they can act as seeds for EW vacuum decay~\cite{Burda:2015isa,Burda:2016mou,Tetradis:2016vqb}, if the EW vacuum is metastable~\cite{Buttazzo:2013uya}. As we have shown, these microscopic BHs are avoided in quadratic gravity if one sets the masses $M_0$ and $M_2$ below $h_{\rm max}$. For central values of the top mass, the Higgs mass and the strong fine structure constant, $h_{\rm max}$ happens to be around $5\times 10^{10}$~GeV~\cite{Espinosa:2015qea}
 and thus is remarkably close to the maximal  $M_2$ required by the Higgs mass naturalness. For  $M_0, M_2\ll h_{\rm max}$ gravity becomes weak for $r_h\lesssim 1/h_{\rm max}$ and the spacetime is a small deformation of the Minkowski background. In this situation gravity cannot have a drastic influence on the EW vacuum decay, unlike what happens in GR.

\section{Light UCO dark matter}
\label{sec:pheno}

Stable UCOs could comprise a fraction or all of DM.  It is, however, not known whether macroscopic gravitationally collapsed objects in quadratic gravity possess a horizon and thus evaporate. We can consider two possibilities:
\begin{itemize}
	\item Gravitational collapse ends in an object with a horizon, in which case the resulting objects can evaporate. Once they have radiated enough mass and their size becomes comparable to $1/M_2$, the horizon will disappear and we are left with a stable relic. Such objects may constitute all DM, as was already noted three decades ago~\cite{MacGibbon:1987my}.
	
	\item A horizon will never form during a gravitational collapse. However, as gravity becomes weak at small distances, gravitational collapse will still end in an object that has a finite size, but is possibly extremely compact. These objects will be stable after their collapse has occurred and, as such, serve as possible DM candidates. Note that when the Schwarzschild radius of the object is larger than $1/M_{0,2}$, then the collapsed object is in the non-linear regime, while when~\eqref{BHcond} is satisfied the linearization mechanism takes place.  In case the UCOs are heavier, this scenario may be experimentally tested via gravitational wave observations of UCO mergers~\cite{Cardoso:2016oxy,Cardoso:2016rao} which are expected to be abundant when they make up a significant fraction of DM ~\cite{Raidal:2017mfl,Ali-Haimoud:2017rtz,Raidal:2018bbj,Vaskonen:2019jpv}.
\end{itemize} 

On the heavier side, above $10^{-12}\Msun$, the abundance of such objects are severely constrained by numerous observables, such as microlensing~\cite{Tisserand:2006zx,Allsman:2000kg,Griest:2013aaa,Zumalacarregui:2017qqd,Garcia-Bellido:2017imq}, gravitational wave observations~\cite{Raidal:2017mfl,Ali-Haimoud:2017rtz,Raidal:2018bbj,Authors:2019qbw,Wang:2016ana,Wang:2019kaf,Vaskonen:2019jpv} or energy injection from accretion~\cite{Ali-Haimoud:2016mbv,Poulin:2017bwe,Hektor:2018qqw,Mena:2019nhm,Hutsi:2019hlw}, to name the most stringent constraints (for recent reviews see e.g. \cite{Carr:2017jsz,Sasaki:2018dmp}). For lighter masses, the abundance for PBHs around $10^{-16}\Msun$ is constrained by the secondary $\gamma$-ray flux form their Hawking radiation, and, at smaller masses, from BBN~\cite{Carr:2009jm}. PBHs lighter than $10^{-17}\Msun$ would have completely evaporated within a Hubble time. Modifications to Hawking radiation can soften the lower bound~\cite{Raidal:2018eoo}. In all, current observations favor lighter objects, with a mass below $10^{-12}\Msun$, and, if they evaporate as BH in GR, they must be lighter than $10^{9} \, \rm g$ to evade interfering with BBN.

Similarly to primordial BHs, UCOs can form in the early universe from the gravitational collapse of large primordial density fluctuations~\cite{Misner:1974qy,Carr:1974nx} which can, for example, be produced during inflation (see e.g.~\cite{GarciaBellido:1996qt,Garcia-Bellido:2017mdw,Kannike:2017bxn,Domcke:2017fix,Ballesteros:2017fsr,Germani:2017bcs,Ezquiaga:2017fvi,Motohashi:2017kbs}). The mass of these objects is smaller than the horizon mass at formation, i.e. $M_{\rm UCO} = \gamma M_{\rm H}$~\cite{Carr:2009jm}, where
\be\label{eq:mH}
	M_{\rm H} 
	= \frac{4\pi}{3}H^{-3}\rho 
	= \frac{4\pi \bp^2}{H}
	= M_{\rm H, eq} \left(\frac{k}{k_{\rm eq}}\right)^{-2}
	= 32 \Msun \left(\frac{k}{\pc^{-1}}\right)^{-2}.
\ee
We assumed that the wavenumber of the collapsing mode is $k = aH$, the radiation domination, i.e. $H\propto a^{-2}$, and used matter-radiation equality for a reference point, $k_{\rm eq} = 0.01 (\Omega_{m}/0.31) \Mpc^{-1}$, $M_{\rm eq} = 3.2 \times 10^{17} \Msun (\Omega_{m}/0.31)^{-2}$~\cite{Green:2004wb}. In GR $\gamma$ is typically $\mathcal{O}(0.1)$ and depends on the size as well on the shape of the density fluctuations (for a recent review see e.g.~\cite{Kalaja:2019uju}). 

Gravitational collapse has not been studied in quadratic gravity, however. For scales $k \ll M_{0,2}$  the effects from the Weyl-squared and $R^2$ terms are suppressed~\cite{Salvio:2019ewf} and the collapse proceeds as in GR. At smaller distances, when $k \gg M_{0,2}$, gravity is weakened and gravitational collapse can be inhibited. To see whether such a scenario might be realized, consider the bound on scale of inflation $H/\bp \leq 
2.7\times 10^{-5}\bp \,\, (95\%~ \mbox{CL})$~\cite{Akrami:2018odb}. This implies that the horizon mass during formation must satisfy $M_{\rm H} \gg 2 \rm g$, because the density perturbations must be produced before inflation ends~\cite{Carr:1994ar}. When the above bound on $H$ is saturated, then the condition for the weakening of gravity $M_{0,2}/H \ll 1$ can be satisfied for $M_{0,2} \approx 10^{-8} \bp$, corresponding to UCOs with a mass $M \lesssim 10^{8} \bp  = 0.4 \rm kg$, in which case critical collapse must be re-examined in the context of quadratic gravity.

The above discussion assumes that the UCOs are cold. If they have a non-vanishing temperature they will lose mass through radiation. In order for a stable remnant to exist, it is necessary that the equation of state of the star supports spherically symmetric configurations at zero temperature. Otherwise, it is possible for the star to lose all of its mass through radiation. This can be achieved, for example, if the matter comprising the star carries a non-vanishing conserved global charge.

The simplest realization of a production mechanism for large primordial perturbations objects is via single field inflation, which tends to favor production of light primordial BHs or UCOs. Eq.~\eqref{eq:mH} shows that lighter objects are produced at smaller scales and thus later during inflation.  Inflationary production of heavier UCOs implies a peak in the power spectrum that was produced not later than 
\be
	\Delta N 
	< \ln(k_{\rm H}/k_{\rm CMB})
	\approx 40 -\frac{1}{2}\ln\left(\frac{M_{\rm H}}{10^{-16} \Msun}\right)
\ee
e-folds after the primordial fluctuations responsible for CMB anisotropies were produced. The relatively short length of the first phase of inflation generally results in a lower value of the spectral tilt of the primordial power spectrum $n_s$ which, for example, for Starobinsky inflation, hilltop inflation or for non-minimally coupled scalars is $n_s \lesssim 1 - 2/\Delta N = 0.95$~\cite{Kannike:2017bxn} and is thus at odds with the observed value $n_s = 0.9649\pm0.0042$~\cite{Akrami:2018odb}. This issue is, however, not present in case DM consists of objects much lighter than $10^{-16}\Msun$ as they may form significantly later.

\section{Conclusions}
\label{sec:end}

Quadratic gravity predicts the existence of light horizonless UCOs that can surpass the compactness of Schwarzschild BHs. This is due to the softening of the gravitational interactions at distances shorter than the Compton wavelengths, $1/M_2$ and $1/M_0$, of the new gravitational degrees of freedom in quadratic gravity. In particular, we show that the weak-field limit can be consistently applied to objects with a Schwarzschild radius smaller than $1/M_2$ and $1/M_0$, even if the size of the objects is much smaller than their Schwarzschild radius. 

By applying the linearized theory, the metric surrounding a general spherically symmetric matter configuration can be obtained by making use of the gravitational propagators of this theory. The resulting spacetimes are regular and asymptotically flat. The results imply simple inequalities which demonstrate the consistency of the weak-field approximation of sufficiently light objects. 

Within the linear theory it is possible to analytically construct UCOs that are more compact than BHs in GR. As concrete examples we considered UCOs consisting of incompressible fluids, which, although they are not completely realistic due to their infinite speed of sound, approximate well non-relativistic fluid stars with a stiff but causal equation of state. Perturbativity of the linear solutions was shown directly by comparing them to numerical solutions in the full-nonlinear theory. We note that the gravitational static spherically symmetric solutions are completely fixed by asymptotic flatness and regularity at the origin, thus, given an equation of state, the configurations are fully determined by a single parameter, e.g. the central pressure. We argued that the solutions can avoid the runaway behavior due to the Ostrogradsky theorem.

We conclude that the absence of a horizon is a general feature of light objects in quadratic gravity. This leads to interesting phenomenological consequences. Moreover, the absence of microscopic BHs improves the consistency of the Standard Model when the EW vacuum is metastable. The UCOs would appear as the endpoint of BH evaporation, if an event horizon can form for heavy objects in quadratic gravity, and in this case they would provide a concrete realization of BH remnants. In any case 
they may serve as candidates of non-particle DM, for example,  UCOs with a mass of about 1 kg are naturally expected for $M_2 \approx 10^{-8} \bp$. Primordial UCO DM can be produced, e.g., via the gravitational collapse of large primordial density fluctuations.

\section*{Acknowledgments}
We thank D.~Anselmi, B.~Assel, L. Buoninfante, A.~Eichhorn, S.~B.~Giddings, M.~Raidal, A.~Strumia  for interesting discussions. This work was supported by the Mobilitas Plus grants MOBTP135 and MOBTT5, the grant IUT23-6 of the Estonian Ministry of Education and Research, and by the EU through the European Regional Development Fund CoE program grant TK133 ``The Dark Side of the Universe'', as well as trough the COST Action CANTATA, supported by COST (European Cooperation in Science and Technology).

\appendix
\section{Reduction of the gravitational field equations}
\label{reduction}

The two independent gravitational field equations can be obtained plugging the metric  \eqref{eq:g_gen} into the gravitational action \eqref{eq:HD_S} and varying with respect to $a$ and $b$. The definition ${\cal G}_{\mu\nu} \equiv 2\kappa g^{-1/2} \delta \mathcal{S} _{\rm G}/\delta g^{\mu\nu}$, where $\mathcal{S} _{\rm G}$ is the gravitational action, implies that
\be
	{\cal G}_{tt} = \frac{2\kappa a^2}{\sqrt{-g}}\frac{\delta \mathcal{S} _{\rm G}}{\delta a}, \qquad
	{\cal G}_{rr} = \frac{2\kappa }{\sqrt{-g}}\frac{\delta \mathcal{S} _{\rm G}}{\delta b},
\ee
The action contains first derivatives of $b$ and second derivatives of $a$. Explicitly 
\be\label{eq:HDterms}
	2\kappa  \mathcal{S} _{\rm G} 
	= \int \td^4 x \sqrt{-g}  \left[ 
		\frac{M_2^2 - M_0^2}{6M_0^2M_2^2}  \frac{b^2}{a^2} {a''}^{2}
	+	\left(\frac{M_2^2 - M_0^2}{6M_0^2M_2^2}  \frac{ba'}{a^2} + \frac{2M_2^2 + M_0^2}{3M_0^2M_2^2} \frac{b}{ra} \right)  b' a''
	+	\ldots
	\right]
\ee
where the ellipsis denotes terms that contain at most first derivatives (or that can be reduced to such terms by adding suitable surface terms). All higher-derivative terms in the equations of motion will, therefore, come from the terms listed above. Let us consider them in detail. Only the ${a''}^{2}$ does generate a term linear in $a^{(4)}$ in ${\cal G}_{tt}$ while terms containing $b'a''$ generate terms linear in $b^{(3)}$ in ${\cal G}_{tt}$. In contrast, terms containing $b'a''$ give rise to terms linear in $a^{(3)}$ in ${\cal G}_{rr}$, while terms containing higher powers of $b'$ (which are not listed above) produce terms linear in $b''$ in ${\cal G}_{rr}$. We remark that when $M_2 \neq M_0$ then the system of equations is also linear in second derivatives of $a$ and $b$.

The terms containing highest derivatives of $a$ in $rr$ and $tt$ field equations \eqref{eq:HD_eom} are thus
 \bea\label{Coeffa4}
 	{\cal G}_{tt} 
&	= \frac{b^2}{3}  \left(\frac{1}{M_0^2}-\frac{1}{M_2 ^2}\right)a^{(4)} + \{ \mbox{ at most 3rd derivatives } \} \, , \\
	{\cal G}_{rr} 
&	=  -\left(\frac{M_2^2 - M_0^2}{6M_0^2M_2^2}  \frac{ba'}{a^2} + \frac{2M_2^2 + M_0^2}{3M_0^2M_2^2} \frac{b}{ra} \right) a^{(3)} + \{ \mbox{ at most 2nd derivatives } \} \,.
\eea
Thus, the derivative of the $rr$ equations contains a term linear in $a^{(4)}$ (with the same coefficient given above for the $a^{(3)}$ term in ${\cal G}_{rr}$) and can be used to eliminate the fourth derivative in the $tt$ equation. This procedure is possible if the coefficient of $a^{(3)}$ in ${\cal G}_{rr}$ is non-vanishing, otherwise the third-order equations may contain singularities. Fortunately, this coefficient cannot be zero for all $r$ because then $a\propto r^{\tau}$, with $\tau\equiv (2M_0^2+4M_2^2)/(M_0^2-M_2^2)$, which we have excluded by regularity either at $r=0$ or at $r\to \infty$.
Once one has eliminated $a^{(4)}$, the system is reduced to two independent third order differential equations, in which  both $a^{(3)}$ and $b^{(3)}$ appear linearly and, furthermore, $b^{(3)}$ is absent from the $rr$ equation. In conclusion, the left hand side of the field equations can be rearranged as
 \bea 
 	{\cal G}_{tt} 
&	= F a^{(3)} + \frac{6 a^2 b}{r \left[\left(M_2^2-M_0^2\right) r a' + 2  \left(M_0^2+2 M_2^2\right)a\right]} b^{(3)} + \{ \mbox{ at most 2nd derivatives } \} \,, \\ 
	{\cal G}_{rr} 
&	= -\left(\frac{M_2^2 - M_0^2}{6M_0^2M_2^2}  \frac{ba'}{a^2} + \frac{2M_2^2 + M_0^2}{3M_0^2M_2^2} \frac{b}{ra} \right) a^{(3)} + \{ \mbox{ at most 2nd derivatives } \} \,,
\eea 
where $F$ is a function of $a$, $a'$, $a''$, $b$, $b'$ and $r$, whose explicit expression is not needed here. Although this procedure involved solving field equations that include sources, these highest derivative terms do not depend on pressure or density.  Finally, note that when $M_0=M_2$ the equation is already of third order and elimination is not needed. 

\section{Bounds for spherically symmetric metric perturbations}
\label{bounds}

To find the upper bound on the metric perturbation we first recast the auxiliary potentials \eqref{eq:VU} from which the metric perturbation can be constructed as
\bea
	 V(r;m) 
	 = - 4\pi G_{N} \int   \td x x^2 &
	 \left( \frac{e^{-mx}}{x} \frac{\sinh(mr)}{mr} \theta(x-r) + \frac{e^{-mr}}{r} \frac{\sinh(mx)}{mx} \theta(r-x)\right) \rho(x) , 
\\	
	 U(r;m) 
	 = - 4\pi G_{N} \int  \td x x^2 &
	 \bigg( \frac{e^{-mx}}{x} (1+mx) \frac{\sinh(mr)}{mr} \theta(x-r) 
\\
	 & \qquad + \frac{e^{-mr}}{r} \frac{\sinh(mx) - mx \cosh(mx)}{mx} \theta(r-x)\bigg) P(x).
\eea
Where we made use of the expression for a shell source \eqref{eq:V_shell}. The limit $m \to 0$ gives
\bea
	 V(r;0) 
&	= - 4\pi G_{N} \int   \td x x^2 
	 \left( \frac{1}{x} \theta(x-r) + \frac{1}{r} \theta(r-x)\right) \rho(x) , 
\\
	 U(r;0) 
&	= - 4\pi G_{N} \int  \td x x^2  \left( \frac{1}{x} \theta(x-r) \right) P(x).
\eea
As expected from the $1/k^4$ asymptotic of the propagators the singularities of the kernels of $V(r;m)$ and $V(r;0)$ cancel. Using the inequalities
\bea
&	0 \leq 1- e^{-x}\frac{\sinh(y)}{y}\leq x,
	\\
&	0 \leq 1- (1+x)e^{-x}\frac{\sinh(y)}{y} \leq \min\left[ x^2/2, x/3 \right],
	\\
&	0 \leq e^{-x} \frac{\sinh(y) - y \cosh(y)}{y} \leq \min\left[ x^2/3, x/6 \right],
	\\
&	0 \leq e^{-x}(1+x) \frac{\sinh(y) - y \cosh(y)}{y} \leq x/2,	
\eea
when $0 \leq y \leq x$, we obtain
\bea
	 |V(r;m) - V(r;0)| \leq 2 G_{N} m \int   \td^3 x |\rho|,
	\\
	 |U(r;m) - U(r;0)| \leq \frac{2}{3} G_{N} m \int   \td^3 x |P|,
	\\
	|r(V'(r;m) - V'(r;0))| \leq \frac{2}{3} G_{N} m \int   \td^3 x |\rho|,
	\\
	|r(U'(r;m) - U'(r;0))| \leq G_{N} m \int   \td^3 x |P|.
\eea
From this \eqref{abLin} follows.

\bibliographystyle{JHEP}
\bibliography{BHinHDG}

\end{document}